\documentclass[12pt,preprint]{aastex}

\slugcomment{submitted to the {\it Astronomical Journal}}

\shorttitle{New High Proper Motion Discoveries}
\shortauthors{Boyd et al.}

\begin{document}

\title{The Solar Neighborhood XXVII: Discovery of New Proper Motion
Stars with $\mu$ $\ge$ 0$\farcs$18 yr$^{-1}$ in the Southern Sky with
16.5 $<$ $R_{59F}$ $\le$ 18.0}

\author{Mark R. Boyd, Todd J. Henry, Wei-Chun Jao} 
\affil{Georgia State University Department of Physics and Astronomy,
Atlanta, GA 30302-4106}
\email{boyd@chara.gsu.edu, thenry@chara.gsu.edu, jao@chara.gsu.edu}

\author{John P. Subasavage}
\affil{Cerro Tololo Inter-American Observatory, La Serena, Chile}
\email{jsubasavage@ctio.noao.edu}

\and

\author{Nigel C. Hambly}
\affil{Scottish Universities Physics Alliance, Institute for
Astronomy, University of Edinburgh, Royal Observatory, Blackford Hill,
Edinburgh EH9 3HJ, Scotland, UK}
\email{nch@roe.ac.uk}

\clearpage

\begin{abstract}

Here we present 1584 new southern proper motion systems with $\mu$
$\ge$ 0$\farcs$18 yr$^{-1}$ and 16.5 $>$ $R_{59F}$ $\ge$ 18.0. This
search complements the six previous SuperCOSMOS-RECONS (SCR) proper
motion searches of the southern sky for stars within the same proper
motion range, but with $R_{59F}$ $\le$ 16.5. As in previous papers, we
present distance estimates for these systems and find that three
systems are estimated to be within 25 pc, including one, SCR
1546-5534, possibly within the RECONS 10 pc horizon at 6.7 pc, making
it the second nearest discovery of the searches. We find 97 white
dwarf candidates with distance estimates between 10 and 120 pc, as
well as 557 cool subdwarf candidates.  The subdwarfs found in this
paper make up nearly half of the subdwarf systems reported from our
SCR searches, and are significantly redder than those discovered thus
far.  The SCR searches have now found 155 red dwarfs estimated to be
within 25 pc, including 10 within 10 pc. In addition, 143 white dwarf
candidates and 1155 cool subdwarf candidates have been discovered.
The 1584 systems reported here augment the sample of 4724 systems
previously discovered in our SCR searches, and imply that additional
systems fainter than $R_{59F}$ $=$ 18.0 are yet to be discovered.

\end{abstract}


\section{Introduction}
\label{sec:intro}

The Research Consortium On Nearby Stars (RECONS)\footnote{\it
www.recons.org} has completed surveying the southern sky for proper
motion objects with $R_{59F}$ (hereafter $R$) brighter than 16.5 mag
using a specialized trawl of the individual plate detections
comprising the SuperCOSMOS Sky Survey database, with new discoveries
dubbed SCR (SuperCOSMOS-RECONS). These results have been detailed in
six papers in the {\it The Solar Neighborhood} (TSN) series:
\cite{2004AJ....128..437H}, \cite{2004AJ....128.2460H},
\cite{2005AJ....129..413S, 2005AJ....130.1658S},
\cite{2007AJ....133.2898F}, and~\cite{2011AJ....142...10B} (TSN VIII,
X, XII, XV, XVIII, and XXV, respectively).  This seventh proper motion
search paper complements the previous six by searching the entire
southern sky for objects with the same lower proper motion cutoff of
$\mu$ $\ge$ 0$\farcs$18 yr$^{-1}$ as in previous papers, while
examining objects fainter than the previous $R$ cutoff.  In this paper
we stretch our search from $R =$ 16.5 to 18.0 to reveal stars
intrinsically fainter than found in previous searches, as well as
stars similar to previous discoveries but at larger distances. In
previous searches, it was found that several stars, including SCR
1845-6357AB, the 23rd nearest system (as of 1 January 2011), had $R$
between 16.0 and 16.5, very near the search cutoff. It is natural,
then, to assume that there may be more very nearby systems just beyond
this cutoff that have heretofore gone unnoticed.

Despite far outnumbering their brighter counterparts, intrinsically
faint nearby stars are underrepresented in the census of nearby stars,
as evidenced by the discovery of many nearby red dwarfs
\citep{2006AJ....132.2360H} and white dwarfs
\citep{2009AJ....137.4547S}.  In particular, of the 256 systems with
accurate parallaxes placing them within 10 pc, 174 (68\%) have
primaries of spectral type M. This search has been designed
specifically to detect such faint objects. This effort is the latest
in a long line of searches that revealed faint nearby stars, such as
the classic sky surveys by
\cite{1971lpms.book.....G,1978LowOB...8...89G} and
\cite{1979lccs.book.....L, 1980PMMin..55....1L}.  Many more surveys
have since been done utilizing modern technology but fundamentally
similar techniques and are discussed in TSN XXV and references
therein.

In this paper, we present photometric distance estimates for the red
dwarf systems found during this search. Using the relations of
\cite{2004AJ....128..437H}, we estimate three red dwarfs to be within
25 pc, including one estimated to be at 6.7 pc. Using reduced proper
motion diagrams, we also identify 97 white dwarf candidates, seven of
which may be within 25 pc (see \S~\ref{sec:rpmd}). Particularly
noteworthy for this survey, we report 557 new cool subdwarf
candidates, nearly doubling the total identified during the SCR
searches. The 1584 systems and 1608 objects reported in this paper
bring the total number of SCR systems to 6308, consisting of 6650
objects\footnote{Note that systems are only counted if all components
are new discoveries. Multiples consisting of two SCR objects are
counted once, while multiples with at least one known component are
not counted towards the number of new systems.}. Identifying
comprehensive samples of proper motion objects now will prepare us for
large-scale astrometric surveys such as LSST and Gaia, as well as
provide precursor investigation opportunities. Ultimately, these new
groups of intrinsically faint objects will help develop accurate
models of the Galaxy as a whole.

\section{Search Criteria and Methodology}
\label{sec:criteria}

This search has been performed with methods similar to those discussed
in previous TSN papers. In brief, we make use of SuperCOSMOS scans of
at least three Schmidt survey photographic plates taken of each field.
The photographic plates scanned into the SuperCOSMOS database are
6$\degr$ $\times$ 6$\degr$ with a 0.5$\degr$ overlap on each side,
giving $\sim$25$\degr$ square of unique sky coverage per field (in
order to streamline computations of astrometric and photometric data,
the overlap regions were not used). For this search, we have examined
893 of 894 fields, with one omitted because of an insufficient epoch
spread and a location too near the Galactic plane. Thus, this search
encompasses the entire southern sky save for one region 25$\degr$
square.

For this search, we require the $R$ magnitude to be between 16.5 and
18.0 mag, with the brighter limit corresponding to the faint limit of
previous searches, and the fainter limit being chosen to reveal late M
dwarfs out to 10 pc. At 25 pc, we estimate that red dwarfs with
spectral types earlier than M7.5V will have apparent $R$ magnitudes
greater than 18.0, with this search specifically targeting M5.5V to
M7.5V.

We have used similar plate ellipticity requirements as TSN XXV: any
object with two or more ellipticities greater than 0.35 is
excluded. This requirement was shown in TSN XXV to eliminate
significant amounts of false detections while removing no real proper
motion objects in a trial sample of $\sim$200 stars. The ellipticity
checks, along with the three-plate detection requirement preserves
high quality candidates that went undetected on a single plate. SCR
1845-6357AB (\cite{2004AJ....128..437H}), for example, was undetected
on the $B$ plate. Finally, we exclude objects with $R_{ESO}$ and $R$
magnitudes differing by more than 1.0 mag in an effort to eliminate
variable giant stars. We find 7626 initial candidates on these plates.

This list of candidates was then checked using Aladin by blinking the
$B_J$, $R$, and $I_{IVN}$ (hereafter $BRI$) plates with a 5$\arcmin$
radius to confirm that the objects are truly moving. 2MASS $JHK_s$
(hereafter $JHK$) magnitudes, coordinates, and observation epochs were
extracted for confirmed proper motion objects that were then checked
against SIMBAD and other proper motion surveys [e.g., NLTT ---
\cite{1980nltt.book.....L}, LEHPM --- \cite{2003A&A...397..575P}, SIPS
--- \cite{2005A&A...435..363D}, UPM --- \cite{2010AJ....140..844F}]
using VizieR to determine which objects are new SCR discoveries. In
both Vizier and SIMBAD, a region 90$\arcsec$ in radius was used to
match candidate objects to published objects, in accordance with the
findings of \cite{2002ApJS..141..187B}, who found that coordinates of
stars in the Luyten half-second (LHS) catalog were usually accurate to
within $\sim$90$\arcsec$ (see their Figure 2). In the blinking
process, several common proper motion (CPM) candidates were noted that
were not extracted in the initial trawl, but were found by eye (see
\S~\ref{sec:cpm}).

\section{Comparison to Previous Searches}
\label{sec:previoius}

Here we compare the discovery statistics of this effort to previous
SCR searches. We continue to use the terminology used in TSN XXV to
describe the various proper motion bins called MOTION, SLOWMO, and
MINIMO, which contain systems with $\mu$ $\ge$ 1$\farcs$00 yr$^{-1}$,
1$\farcs$00 yr$^{-1}$ $>$ $\mu$ $\ge$ 0$\farcs$50 yr$^{-1}$, and
0$\farcs$50 yr$^{-1}$ $>$ $\mu$ $\ge$ 0$\farcs$18 yr$^{-1}$,
respectively. The two lower cutoffs were chosen to match those of the
Luyten Half Second (LHS) and Luyten Two Tenths (LTT) catalogs,
respectively. The results are summarized in Table~\ref{discostats},
where the first number in each column is from previous searches while
the second is from this effort. We note that significant amounts of
garbage (i.e., false detections) were found in this search. These
candidates, as in previous searches in high proper motion regimes, are
primarily detections with erroneously high proper motion
data. Matching high proper motion objects across epochs is more prone
to mismatches than matching slow-moving or non-moving objects,
particularly at faint magnitudes, hence the amount of garbage at
higher proper motions is significantly larger.

We also update the hit rates --- the sums of new, known, and duplicate
objects divided by the total starting sample of all candidate objects
--- for each proper motion bin.  Known objects are defined as those
found to be previously discovered by other groups as outlined in
\S~\ref{sec:criteria}, whereas duplicates are objects for which more
than one hit was returned from the SuperCOSMOS database.  Garbage
objects are those that are returned from the search but were found to
be erroneous point source matches that were not verified to be proper
motion objects.  The new hit rates for the SCR sample are 7.1\%,
81.0\%, and 84.6\% for the MOTION, SLOWMO, and MINIMO samples,
respectively. The MOTION hit rate is slightly lower than the most
recent update in TSN XXV due to the substantial amounts of garbage
found in this search, while hit rates for the other two samples are
somewhat higher, due to search criteria excluding excessive garbage in
those proper motion bins.

In the NLTT catalog, there are 4356 objects that meet the criteria of
this search. Of these, we recover 2782, or 63\%, which is somewhat
lower than the 71\% recovered during the search in TSN XXV.  Our
cumulative total of 6308 systems for our entire southern hemisphere
search represents an increase of 24.6\% over the 25616 entries in the
southern NLTT catalog.

\section{Data}
\label{sec:data}
  
In Table~\ref{faintmfive}, we list the first five of the 1608 newly
discovered objects. All 1608 are listed in the electronic version of
{\it Astronomical Journal}. We provide SCR names, J2000.0 coordinates,
SuperCOSMOS proper motions and position angles, $BRIJHK$ magnitudes,
the $(R-J)$ color, and a distance estimate. We report a total of 1584
systems, which is less than the number of objects because (a) some
systems have more than one SCR object and (b) systems including both a
known and SCR object are not included in the number of systems. We
list red dwarfs within 25 pc, white dwarfs, and cool subdwarfs in
Tables~\ref{25}, ~\ref{wd}, and~\ref{sdfive}, respectively. The white
dwarfs and subdwarfs have been selected according to the criteria
outlined in \S~\ref{sec:rpmd}. Multiple systems are listen in
Table~\ref{cpm} and discussed in \S~\ref{sec:cpm}. 

TSN XVIII and XXV, as well as~\cite{2010AJ....140..844F}, compare
SuperCOSMOS data to data from various catalogs such as {\it
Hipparcos}, Tycho, PPMX, and PPMXL. The results indicate that the SCR
proper motions are consistent to other sources to $\sim$0$\farcs$02
yr$^{-1}$in both RA and DEC. The details of those comparisons can be
found in the TSN XVIII and XXV.

The distance estimates in
Tables~\ref{faintmfive},~\ref{25},~\ref{sdfive}, and~\ref{cpm} have
been derived using the suite of 11 plate relations in
\cite{2004AJ....128..437H}, which utilizes the six photometry values,
--- $BRI$ plate magnitudes from SuperCOSMOS and $JHK$ from 2MASS ---
associated with each object. That paper and TSN XXV describe errors
associated with the relations in more detail. The plate photometry
errors are $\sim$0.3 mag, while 2MASS data errors are typically less
than $\sim$0.03 mag. The relations fail for blue stars, as well as
white dwarfs and most subdwarfs. For this reason, we list such
objects' distances in brackets in the
Tables~\ref{faintmfive},~\ref{25},~\ref{sdfive}, and~\ref{cpm}. White
dwarfs have more accurate distances listed in the notes of
Tables~\ref{faintmfive},~\ref{wd}, and~\ref{cpm}.

\section{Analysis}
\label{sec:analysis}

In this search we have found a few red dwarfs estimated to be within
25 pc, as well as many white dwarf and cool subdwarf candidates. The
numbers in these categories, however, are significantly different from
previous papers. We find only three red dwarfs within 25 pc, listed in
Table~\ref{25}. We estimate the nearest, SCR 1546-5534, to be at 6.7
pc. If this estimate holds true, it will become the second nearest
object found by the SCR searches, after SCR 1845-6357AB at 3.5 pc
(\cite{2006AJ....132.2360H}). We find a total of 97 white dwarfs using
two reduced proper motion diagrams (see \S~\ref{sec:rpmd}), tripling
the total number of candidates from previous SCR searches, during
which we found 46. Seven of the candidates from this search are
estimated to be within 25 pc. Of particular note, we find 557 new cool
subdwarf candidates, many of which are redder than those previously
found. This total nearly doubles the 598 found in all previous SCR
searches and comprises over one third of the sample in this paper. The
remaining objects are likely to be early- to mid-type M dwarfs beyond
25 pc.

\subsection{Color-Magnitude Diagram}
\label{sec:cmd}

\begin{figure}
\centering
\includegraphics[angle = 90, scale = 0.75]{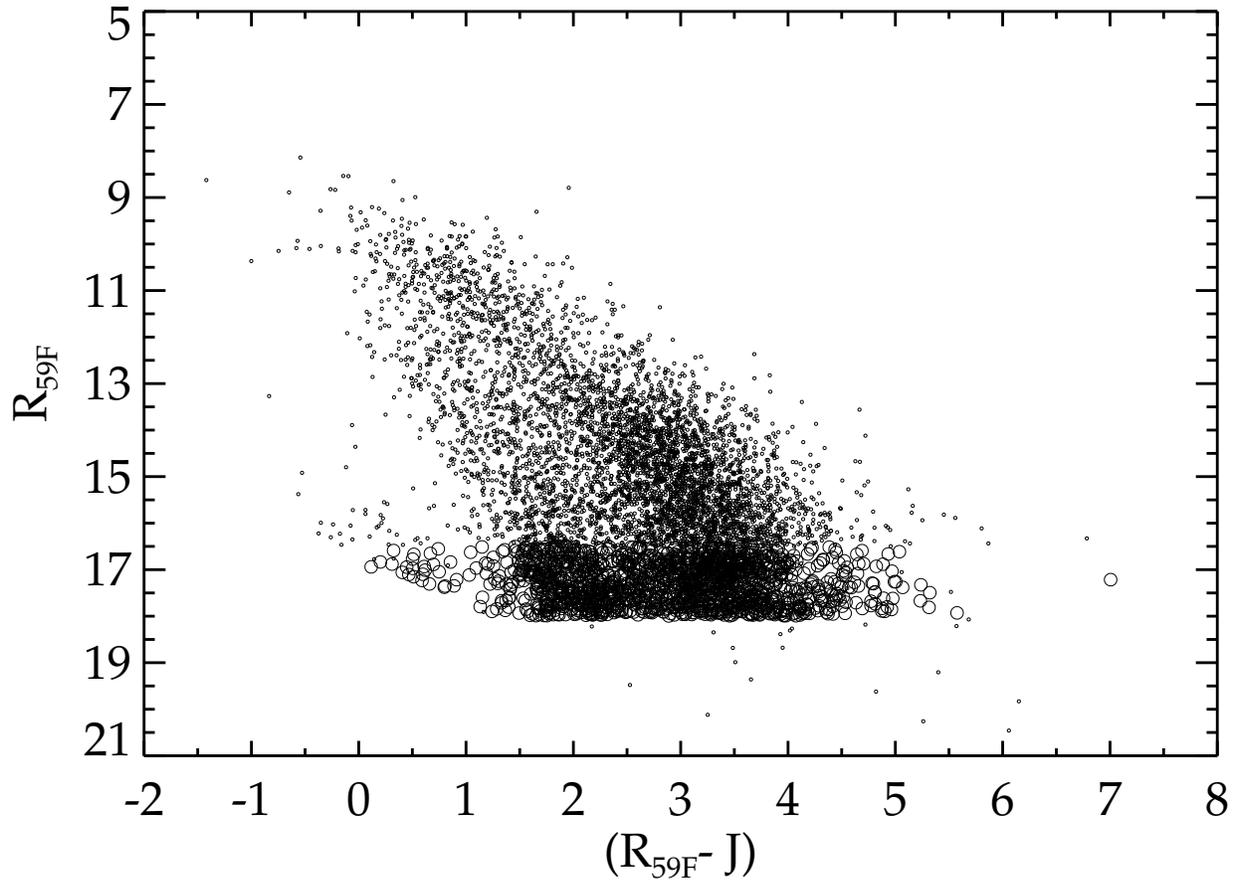}
\caption{Color-apparent magnitude plot for all SCR
discoveries. Discoveries from this paper are marked with open circles,
while those from previous papers are marked with small
circles. Scattered points below $R_{59F}$ $=$ 16.5 are CPM companions
noticed by eye}
\label{cmd}
\end{figure}

Figure~\ref{cmd} shows the color-magnitude diagram for all SCR
discoveries to date. Discoveries from the previous six searches are
marked with small circles, while those presented here are marked with
large circles. This search reveals sources that are redder than
previous searches. Of note is the point at $(R-J)$ $=$ 7.01, which is
SCR 1546-5534, the nearest and reddest system discovered by this
search. The bluest objects are all white dwarf candidates, with the
bluest being SCR 0221-0143, with $(R-J)$ $=$ 0.12.

While not as easily discerned as in previous papers, the white dwarf
and subdwarf regions are visible, with the white dwarfs extending to
the left of $(R-J)$ $=$ 1.0 and the subdwarfs found between $(R-J)$
$=$ 1.0 and $(R-J)$ $\sim$ 2.5. Further discussion of those systems
can be found in the next three sections.

\subsection{Reduced Proper Motion Diagram}
\label{sec:rpmd}

\begin{figure}
\centering
\includegraphics[angle = 90, scale = 0.75]{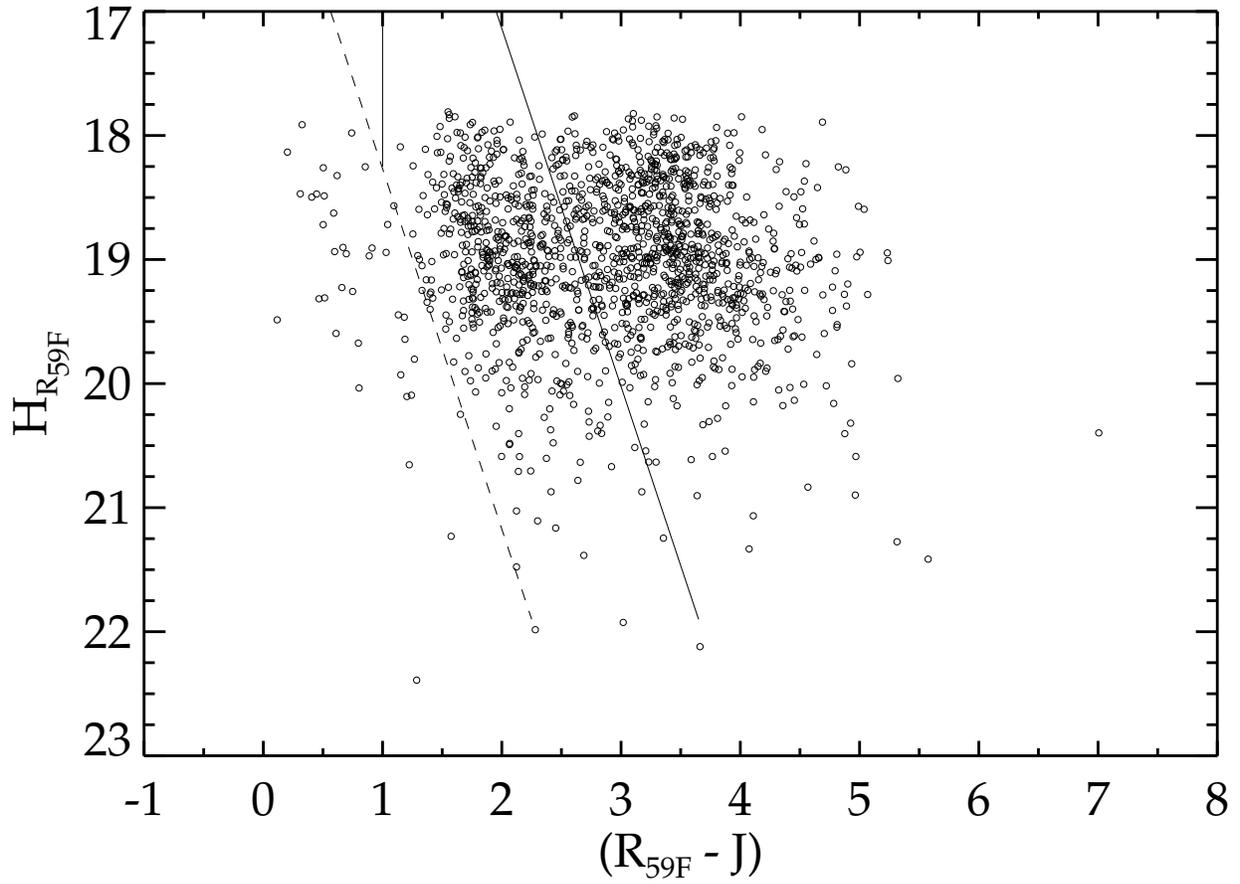}
\caption{$(R-J)$ reduced proper motion diagram. The dashed line
separates white dwarfs from subdwarfs, while the solid lines, along
with the dashed line, outline the cool subdwarf region.}
\label{rpmd1}
\end{figure}

\begin{figure}
\centering
\includegraphics[angle = 90, scale = 0.75]{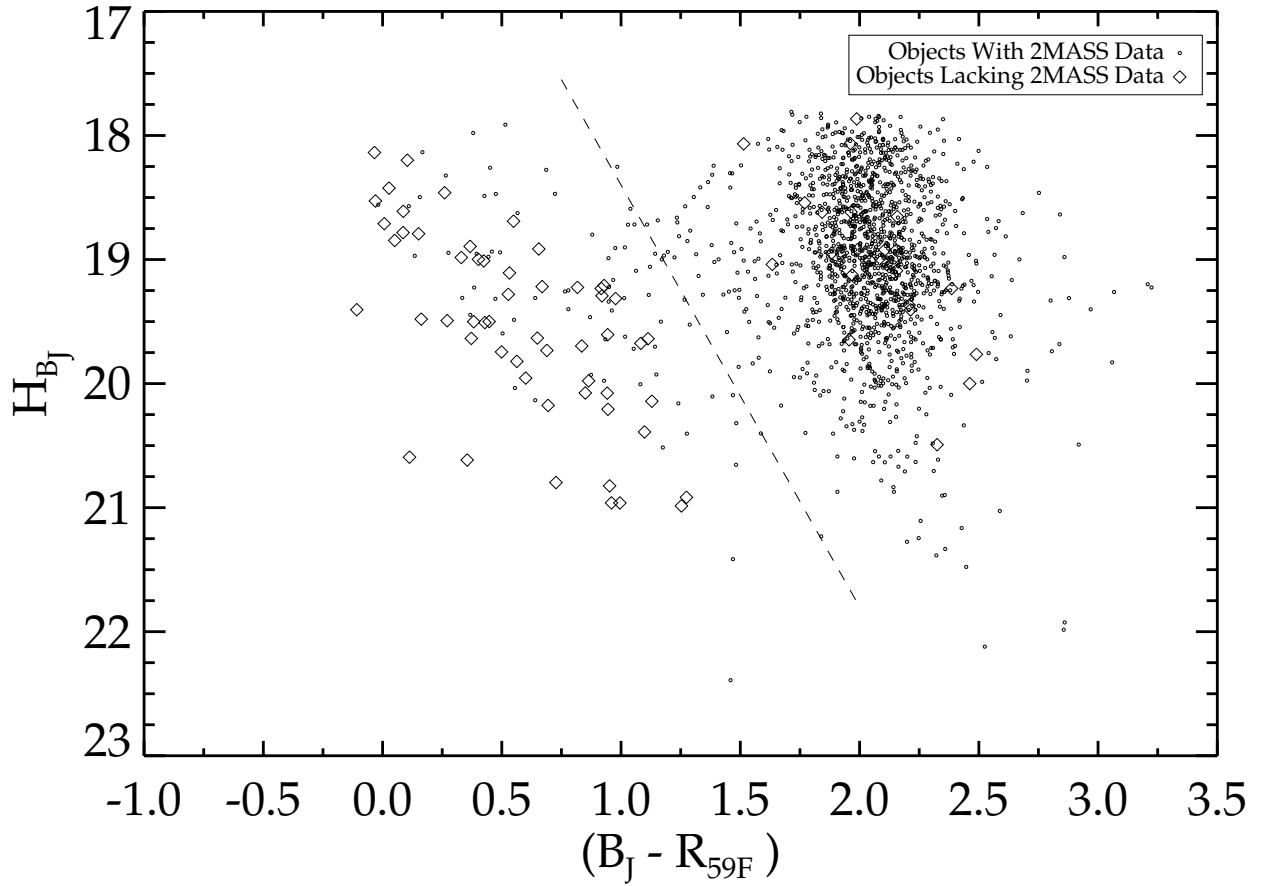}
\caption{$(B-R)$ reduced proper motion diagram. The dashed line
separates white dwarfs from subdwarfs and red dwarfs. Objects lacking
2MASS data are marked with diamonds. All such objects below the dashed
line have been selected as WD candidates.}
\label{rpmd2}
\end{figure}

Figures~\ref{rpmd1} and~\ref{rpmd2} show reduced proper motion (RPM)
diagrams for SCR objects found during the present survey.  The RPM
diagram is a powerful tool for estimating the luminosity class of a
star. It is similar in nature to the H-R diagram except that the
distance is replaced by the star's proper motion, relying on the
inverse statistical relationship between proper motion, $\mu$, and
distance. While obviously not foolproof --- for example, white dwarfs
may mimic subdwarfs, and vice versa in the RPM diagram --- the diagram
allows for the rough classification of systems.  The equation used
here to determine the pseudo-absolute magnitude $H_{R_{59F}}$ plotted
on the vertical axis of Figure~\ref{rpmd1} is

\begin{displaymath}
H_{R_{59F}} = R_{59F} + 5 + 5\log\mu.
\end{displaymath}

The dashed line used to separate white dwarf and subdwarf candidates
in Figure~\ref{rpmd1} is the same as in RPM diagrams in previous
papers. The solid lines, as in TSN 25, denote the cool subdwarf
region. Cool subdwarfs are defined as having $(R-J)$ $>$ 1.0 and are
within four magnitudes of the white dwarf line. These definitions of
both the white dwarf and subdwarf regions are, of course, arbitrary,
but have proven to be reliable in selecting high quality
candidates. We find a total of 42 white dwarf candidates and 557 cool
subdwarf candidates in this plot.

New to our SCR searches, we have included a second RPM diagram in
addition to the $(R-J)$ version. To reveal white dwarfs too faint to
be in 2MASS, we use the $(B-R)$ color because many WDs will not show
up in 2MASS given our search limits of 16.5 $<$ $R$ $\le$ 18.0 and the
typical $(R-J)$ $<$ 1.0 color of WDs. In order to calculate $H_{B}$ we
use the equation above, substituting $B$ for $R$. We use the same
WD-subdwarf cutoff line as \cite{2001Sci...292..698O} to identify an
additional 55 white dwarf candidates. We do not select any more cool
subdwarf candidates because subdwarfs and main sequence stars are
indistinguishable in this RPM diagram. Further discussion of white
dwarfs and subdwarfs can be found in Sections~\ref{sec:wd}
and~\ref{sec:sd}.

\subsection{New White Dwarf Candidates}
\label{sec:wd}

We find a total of 97 new white dwarf candidates using the two RPM
diagrams, presented in Table~\ref{wd} and divided into two categories
based on the RPM diagram from which they were selected. We present the
same data as Table~\ref{faintmfive} except in lieu of the inaccurate
distance estimates generated assuming the objects are red dwarfs, we
give distances based on the single color linear relation of
\cite{2001Sci...292..698O}, assuming the objects are white dwarfs. We
find seven WDs estimated to be within 25 pc, all from the $(R-J)$ RPM
diagram. Of these, five have either optical or IR colors that are
redder than typical white dwarfs. Two, SCR 0058-6905 and SCR
0816-4641, do appear to be good nearby white dwarf candidates. Three
of the seven 25 pc candidates presented lie within the 13 pc horizon
within which \cite{2002ApJ...571..512H} suggested the white dwarf
sample is largely complete. All three have colors redder than typical
WDs, suggesting they may in fact be subdwarfs contaminating the white
dwarf region of the RPM diagram. 

The faintness of the WDs selected via the $(B-R)$ RPM diagram suggests
they are all quite distant, and indeed the distance estimates
according to the relation of \cite{2001Sci...292..698O} show the
nearest to be at 40 pc.

TSN XXV discusses the 46 previous WD candidate discoveries, while the
current effort adds 97 objects to this sample. In total we estimate 17
to be within 25 pc, of which seven are from this paper. TSN XXV
discusses parallax results from CTIOPI for four of these white dwarfs,
confirming they are within 25 pc. Spectroscopic observations are
underway for unconfirmed WD discoveries.

\subsection{New Cool Subdwarf Candidates}
\label{sec:sd}

\begin{figure}
\centering
\includegraphics[angle = 90, scale = 0.75]{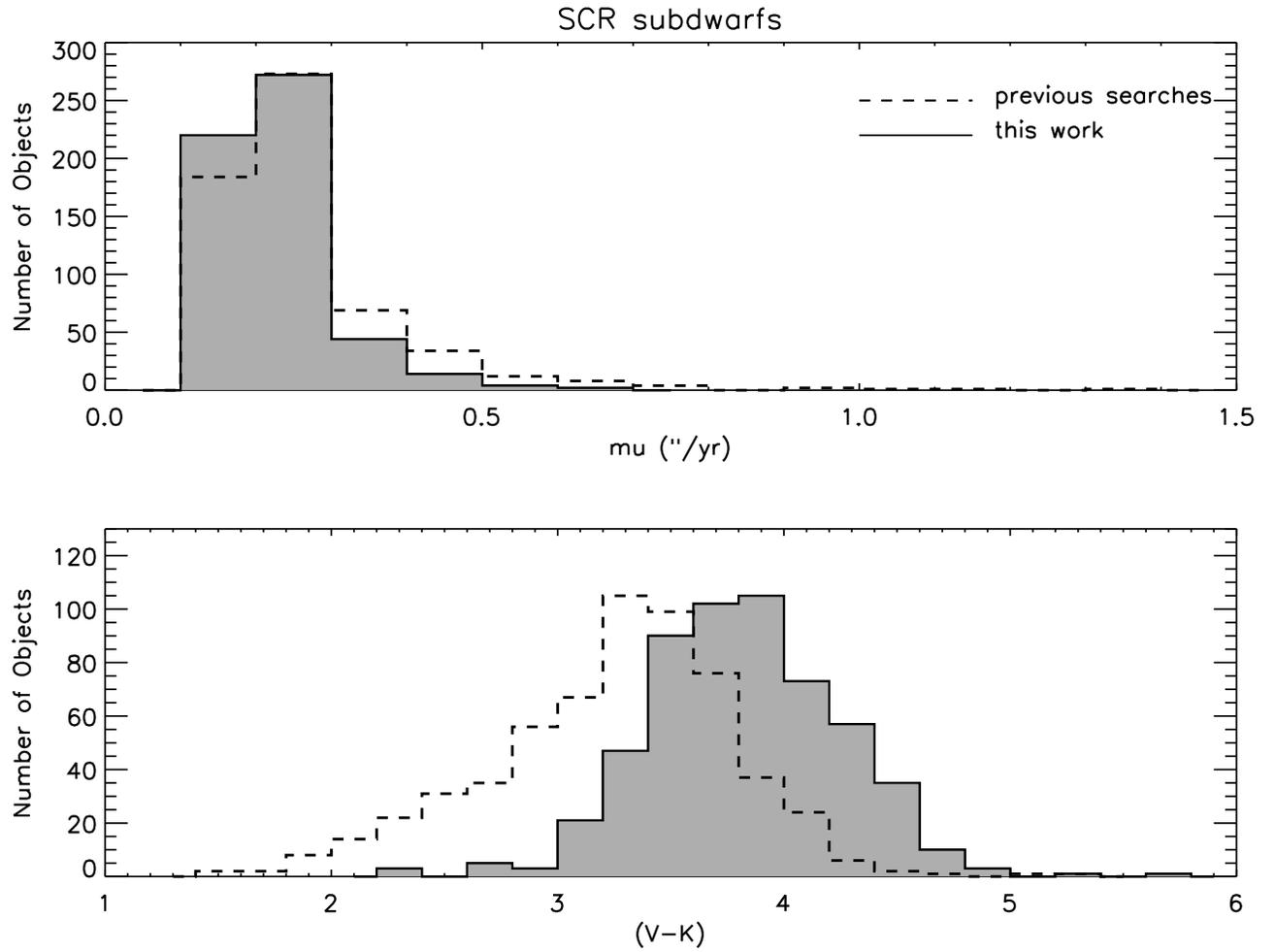}
\caption{Histograms of SCR subdwarf candidates this paper (filled) and
previous TSN papers (outlined). The top plot shows quantity versus
proper motion bin, while the bottom shows quantity versus color. There
are notably fewer high proper motion discoveries in this paper, but
those subdwarf candidates discovered tend to be redder than previous
discoveries.}
\label{sdcomp}
\end{figure}

Table~\ref{sdfive} provides the first five of the 557 cool subdwarf
candidates.  Because of their faintness compared to their
similarly-colored main sequence counterparts, the distance estimates
reported are too large and are listed in brackets in
Tables~\ref{faintmfive} and~\ref{sdfive}.  Figure~\ref{sdcomp} shows
histograms of subdwarfs found in previous SCR searches as well as this
work. The top plot shows we find more subdwarfs in the bin from
0$\farcs$18 yr$^{-1}$ $\le$ $\mu$ $\le$ 0$\farcs$20 yr$^{-1}$, but
fewer subdwarfs with $\mu$ $>$ 0$\farcs$30 yr$^{-1}$ than previous
works. The bottom plot shows we discover redder subdwarfs in this
work. The mean $(V-K)$ color\footnote{The $V$ magnitudes are estimated
using the mean of the $B$ and $R$ magnitudes.} of subdwarfs discovered
in previous efforts is 3.2, while it is 3.8 in this effort.  The
redness may be explained in part by these systems being more distant
discoveries than those from previous papers, and hence being affected
by interstellar reddening.  In the previous searches, the reddest
subdwarf candidate has $(V-K)$ $\sim$ 5.21 (SCR 0020-2642B), while
here we find one with $(V-K)$ $\sim$ 5.74 (SCR 1522-0244), which is
similar in color to LHS 377, an M7.0 subdwarf
(\cite{1997AJ....113..806G}, $(V-K)$ $=$ 5.76). The HR diagrams
in~\cite{1997AJ....113..806G} and~\cite{2008AJ....136..840J} show that
there are very few cool subdwarfs redder than M5.0. In recent years a
few L type subdwarfs have been identified (\cite{2009AIPC.1094..242B}
and references therein), but more are needed to to explore the
hydrogen burning limit that comprise the end of main sequence.  This
work has increased the number of cool subdwarf candidates, but future
astrometric and spectroscopic observations are needed to confirm them.

\subsection{Common Proper Motion Systems}
\label{sec:cpm}

We have found 24 potential common proper motion (CPM) systems,
including one triple, listed in Table~\ref{cpm}. Of these, 19 consist
entirely of new discoveries, while the other five have at least one
known component. In Table~\ref{cpm} we provide distance
estimates for all components, proper motions, separations, and
position angles for both proper motions and separations.

Systems were found by one of two methods. First, during the blinking
process systems were noted as having similar proper motion and
position angles. Second, an automated search for companions within a
radius of 1200$\arcsec$, as well as $\Delta$$\mu$ $<$ 0$\farcs$025
yr$^{-1}$ and $\Delta$ PA $<$ 10$\degr$, was performed around each
object in the search.

The database from which the initial sample was drawn is a version of
the database optimized for high proper motion objects.  For systems
noticed by eye that were not in the initial search sample, we manually
gathered SuperCOSMOS data from the SuperCOSMOS Sky Survey (SSS)
website. For moving objects, the SSS online catalogs tend to be less
reliable and less complete, even at relatively low proper motions, as
discussed in TSN XXV.  We divide the table into two sections, probable
and possible systems. Probable systems have $\mu$ and PA within
0$\farcs$025 yr$^{-1}$ and 10$\degr$, respectively, as well as
consistent distance estimates, while probable systems appear to be
comoving on plate images but either lack data to confirm common proper
motion or have data that suggest otherwise.

\begin{figure}
\centering
\includegraphics[angle = 90, scale = 1]{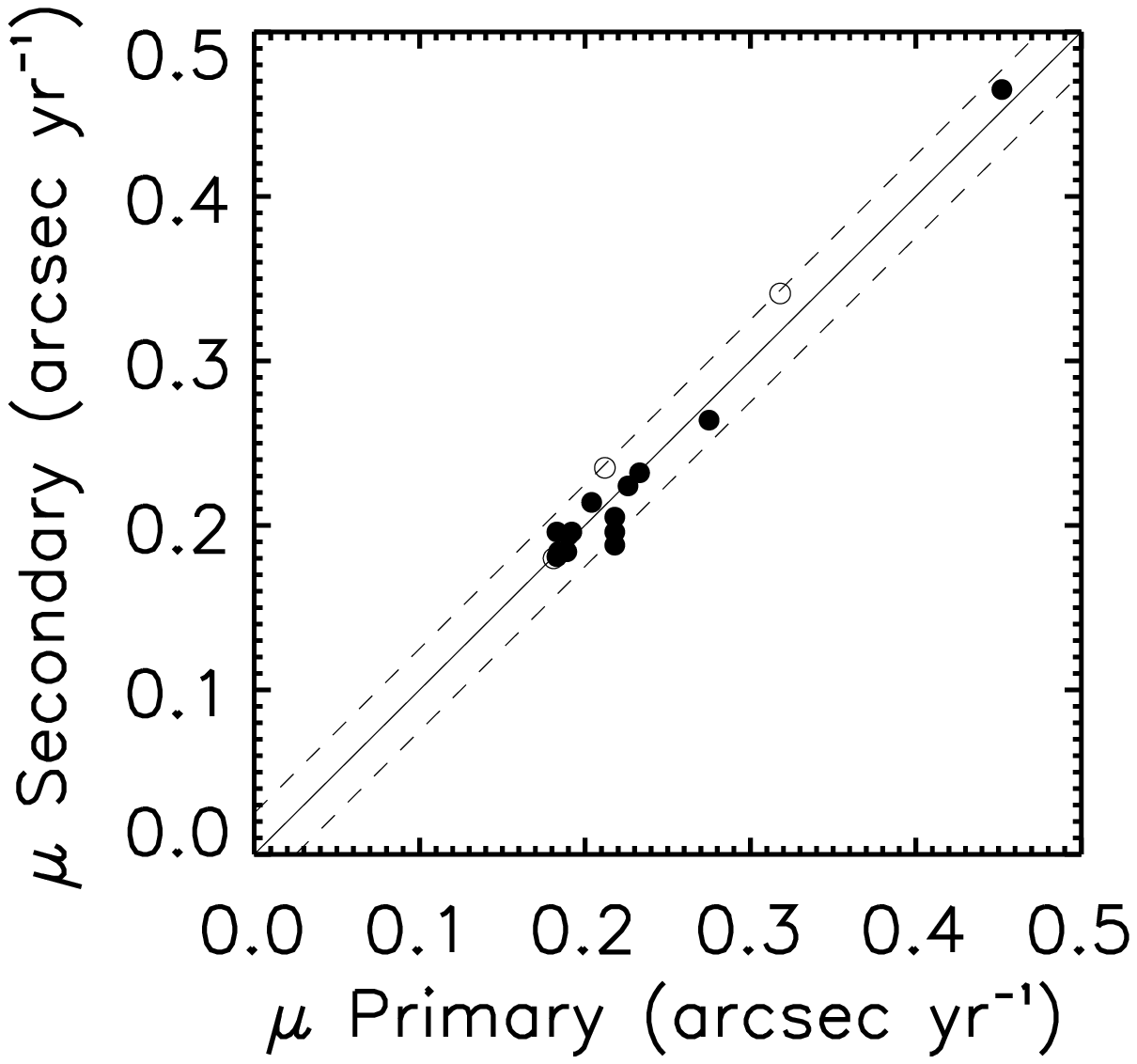}
\caption{Plot of the proper motion of the primary versus that of its
companion in common proper motion systems. The solid line denotes
perfect agreement between the two proper motions, while the dashed
lines indicate limits of 0$\farcs$025 yr$^{-1}$ in accordance with our
uncertainties. Filled circles represent pairs in which both members
had data from the automatic phase of the search. Open circles denote
pairs in which proper motion data for at least one component were
gathered manually.}
\label{mumu}
\end{figure}

\begin{figure}
\centering
\includegraphics[angle = 90, scale = 1]{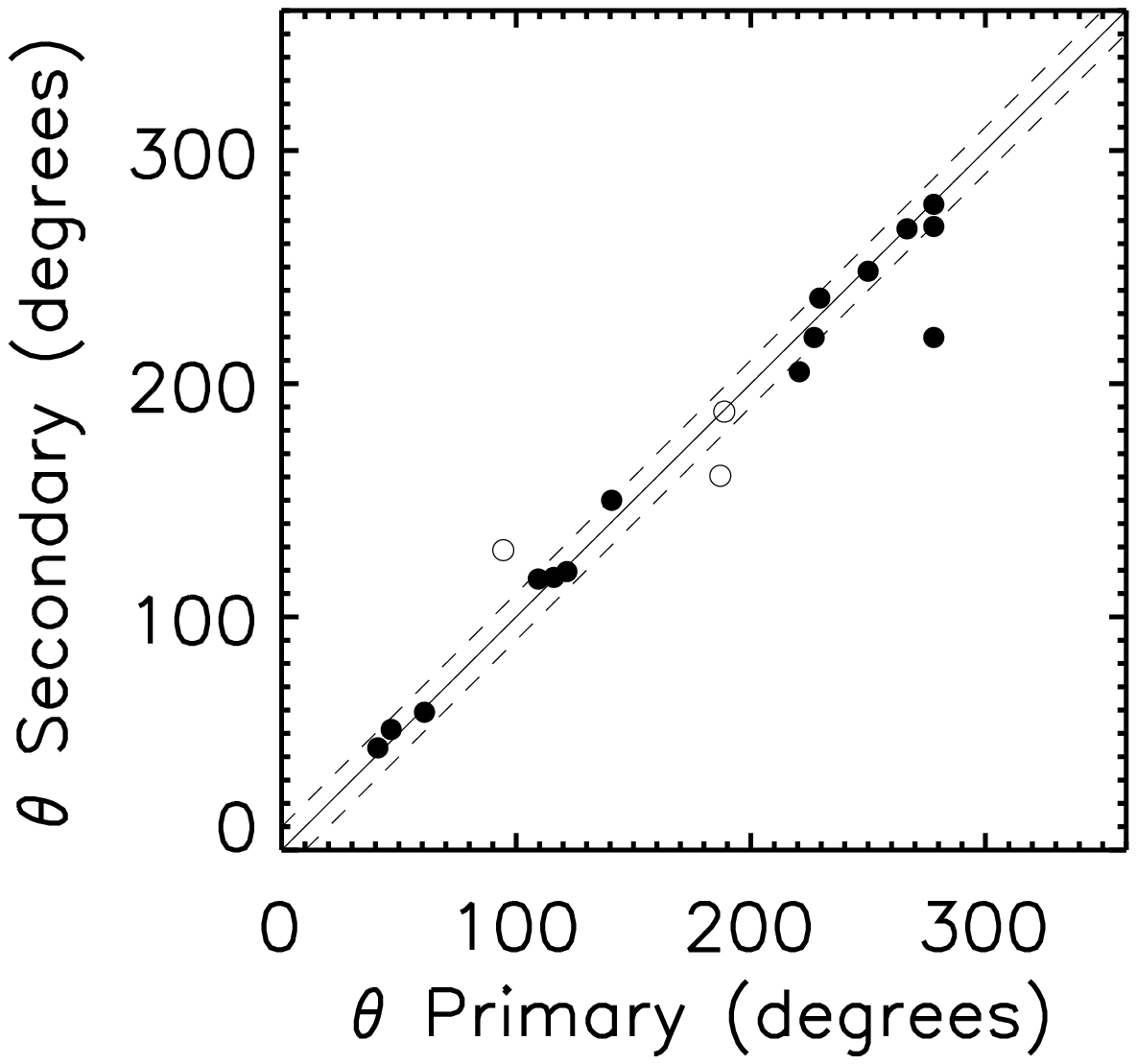}
\caption{Plot of the position angle of the proper motion of the
primary versus that of its companion in common proper motion
systems. The solid line denotes perfect agreement between the two
proper motions, while the dashed lines indicate limits of 10$\degr$in
accordance with our uncertainties. Filled circles represent pairs in
which both members had data from the automatic phase of the
search. Open circles denote pairs in which proper motion data for at
least one component were gathered manually.}
\label{papa}
\end{figure}

Figures~\ref{mumu} and~\ref{papa} compare the proper motion sizes and
position angles for components of multiple systems,
respectively. Systems marked with solid points are those with both
components retrieved during the initial automated search, and tend to
have better agreement between these values, particularly for the
proper motions. Systems for which at least one component had its
proper motion and position angle data gathered manually from the SSS
database are marked with open circles.

\subsection{Sky Distribution of SCR Systems}
\label{sec:allscr-distro}

\begin{figure}
\centering
\includegraphics[angle = 90, scale = 0.75]{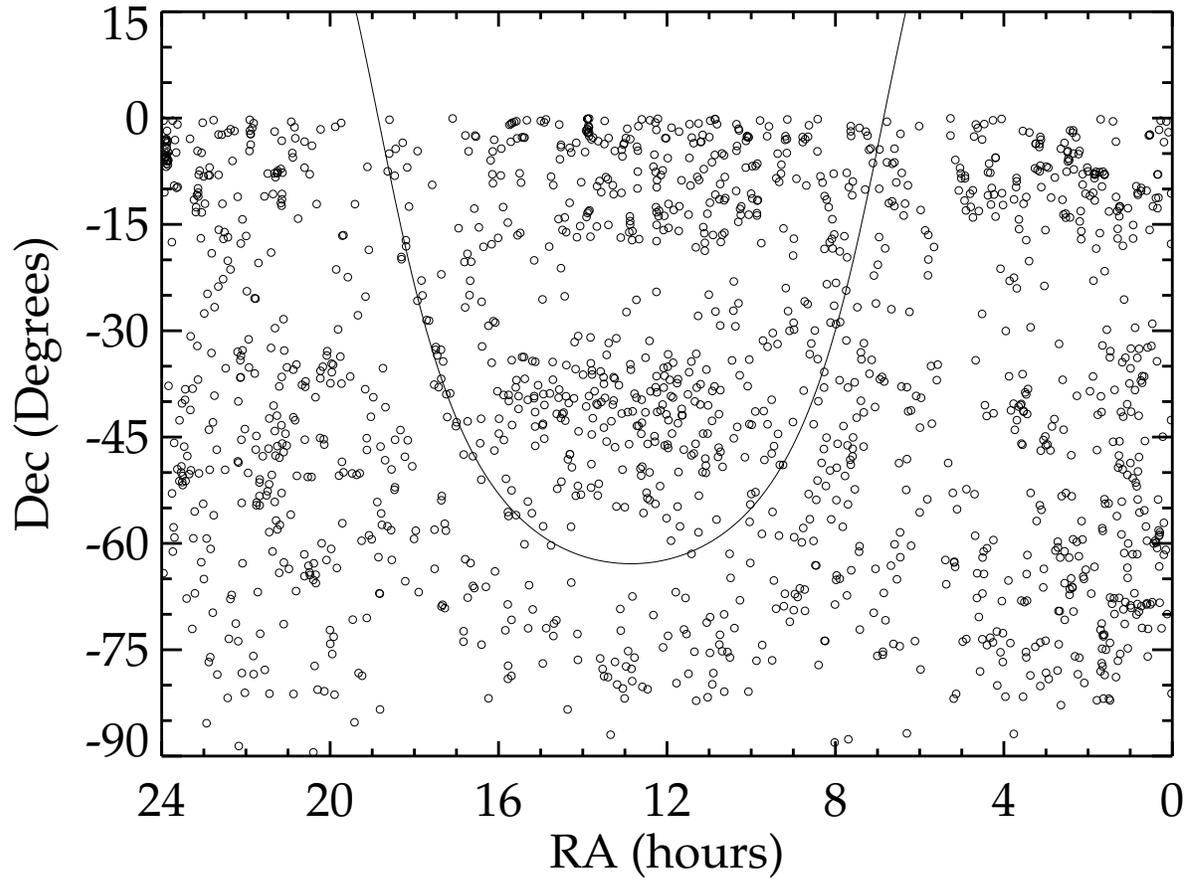}
\caption{Sky distribution of newly discovered systems. The curve
represents the Galactic plane. }
\label{distro}
\end{figure}

Figure~\ref{distro} shows the sky distribution of systems presented in
this paper, with the curve representing the Galactic plane. As in TSN
XXV, we note a string of new systems superimposed on background dust
in the Galactic plane. We also note a lack of discoveries in the areas
near the Galactic center not obscured by dust. When compared to Figure
7 in TSN XXV, the NLTT system distribution in the southern sky, we
note a similarly high density of new systems in areas searched less
thoroughly by Giclas and Luyten. In the area between declinations
-15$\degr$ and -30$\degr$, there are relatively few new systems. This
is because this region of the sky has been searched particularly
deeply, notably by Giclas and Luyten as well as the LEHPM survey. A
similar pattern exists, but was not noted, in the distribution
presented in Figure 6 of TSN XXV.

\subsection{Comments on Individual Systems}
\label{sec:individual}

Here we highlight a few of the 1584 systems reported from this portion
of the SCR search.  These represent extremes in proximity, color (both
blue and red), and/or sky separation.

{\bf SCR 0058-6905} is the second nearest white dwarf candidate,
estimated to be at 12.1 pc, and one of three within the 13 pc limit of
completeness according to \cite{2002ApJ...571..512H}. Its nature as a
white dwarf is somewhat questionable, however, as its $(B-R)$ color is
redder than most white dwarfs.

{\bf SCR 0128-7401} is the third nearest white dwarf candidate,
estimated to be at 12.8 pc, and one of three within the 13 pc limit of
completeness according to \cite{2002ApJ...571..512H}. Its optical and
near IR colors, however, are redder than most white dwarfs, calling in
to question its nature.

{\bf SCR 0221-0143} is the bluest system discovered in this search
with $(R-J)$ $=$ 0.12. It is a white dwarf candidate at 46.5 pc.

{\bf SCR 0307-4654A/LEHPM 1-3098} is the most widely separated system
in this search, with a separation of 1180$\farcs$3. Using the mean
distance of the system, this gives a projected sky separation of
nearly 200,000 AU. While this system may be a random alignment of two
unrelated stars, if the two stars comprise a system, it would be among
the most weakly bound and widest systems known.

{\bf SCR 1546-5534} is the nearest star revealed by this search. It is
the second closest system revealed by the SCR searches, after SCR
1845-6357AB at 3.5 pc, as reported in \cite{2006AJ....132.2360H}. With
$(R-J)$ $=$ 7.01, it is also the reddest object revealed by the SCR
searches. We note that this object only has eight plate relations,
with distance estimates ranging from 3.1 to 15.7 pc, calling into
question its estimate of 6.7 pc. Even so, it is worthy of follow up
and is currently being observed as part of RECONS' ongoing parallax
program at the CTIO 0.9m telescope.

{\bf SCR 2045-1411} is the nearest estimated white dwarf candidate at
10.7 pc, and one of three that lie within the 13 pc limit of
completeness according to \cite{2002ApJ...571..512H}. It is, however,
unlikely to be an actual WD. Its $(B-I)$ and $(J-K)$ colors are
significantly redder than known white dwarfs, and unlike such known
white dwarfs, it satisfies all 11 red dwarf plate relations. It is
likely a cool subdwarf candidate contaminating the white dwarf area of
the RPM diagram.

\section{Discussion}
\label{sec:discussion}

This paper completes an SCR sweep of the southern sky to $R$ $=$ 18.0
for systems with $\mu$ $\ge$ 0$\farcs$18 yr$^{-1}$. In total, we have
found 6308 new proper motion systems, of which 1584 are first reported
in this paper, representing an increase of 33\% over the previous
count. A total of 155 red dwarfs estimated to be within 25 pc,
including 10 within 10 pc, have been found, of which three and one
originate from this paper, respectively. We have found an additional
97 white dwarf candidates, bringing the total to 143, and an
additional 557 cool subdwarf candidates, bringing the total to 1155.

This paper is unique for the sheer numbers of white dwarf and subdwarf
candidates revealed, with total counts that have tripled and doubled,
respectively. Given the nature of their selection, there is obviously
some contamination of the samples, but regardless, this represents a
substantial increase. Of the 97 white dwarfs presented here, seven are
estimated to lie within 25 pc. The subdwarfs presented here are redder
and cooler than those previously discovered, helping fill in a
currently underrepresented sample of cool subdwarfs discovered by
these searches. 

As part of our Cerro Tololo Inter-American Observatory Parallax
Investigation (CTIOPI), an astrometry program carried out at the CTIO
0.9m [\cite{{2005AJ....129.1954J}}, \cite{2006AJ....132.2360H},
\cite{2009AJ....137.4547S}, \cite{2010AJ....140..897R}, and Jao et
al. (2011)], we aim to obtain accurate trigonometric parallaxes for
systems estimated to be near the Sun from the categories of 25 pc
objects, white dwarfs, and cool subdwarfs. While the sheer number of
candidates for observation means we can only select the best gems, the
SCR search efforts will provide a wealth of targets for particular
attention in large scale astrometric efforts such as Gaia and
LSST. Such observations will help us more map the solar neighborhood
more accurately than ever before.

\section{Acknowledgments}

The RECONS effort is supported by the National Science Foundation
through grant AST 09-08402. Funding for the SuperCOSMOS Sky Survey was
provided by the UK Particle Physics and Astronomy Research Council.
N.C.H.~would like to thank colleagues in the Wide Field Astronomy Unit
at Edinburgh for their work in making the SSS possible; particular
thanks go to Mike Read, Sue Tritton, and Harvey MacGillivray.  This
research has made use of results from the SAO/NASA Astrophysics Data
System Bibliographic Services, the SIMBAD and VizieR databases
operated at CDS, Strasbourg, France, and the Two Micron All Sky
Survey, which is a joint project of the University of Massachusetts
and the Infrared Processing and Analysis Center, funded by NASA and
NSF.

\clearpage


\clearpage


\hoffset-00pt{}
\begin{deluxetable}{lcccr}
\tablecaption{Discovery Statistics for Entire SCR Sample to Date\tablenotemark{a}
\label{discostats}}
\tablewidth{0pt}

\tablehead{\vspace{-8pt} \\
\colhead{Category}&
\colhead{MOTION\tablenotemark{b}}&
\colhead{SLOWMO\tablenotemark{c}}&
\colhead{MINIMO\tablenotemark{d}}&
\colhead{Total\tablenotemark{e}}}

\startdata

New Discoveries      &    9 $+$ 0  &  141 $+$ 9  &  4574 $+$ 1575&  6308 \\ 
Known                &  171 $+$ 11 & 1159 $+$ 133& 17244 $+$ 4090& 22808 \\ %
Duplicates           &   15 $+$ 1  &   91 $+$ 3  &  1640 $+$ 356 &  2106 \\ %
Garbage              & 1989 $+$ 700&  344 $+$ 17 &  5335 $+$ 28  &  8413 \\ %
Total hits           & 2184 $+$ 712& 1735 $+$ 162& 28793 $+$ 6049& 39635 \\ %

\enddata

\tablenotetext{a}{A few objects were also reported in searches done concurrently by
\cite{2005A&A...435..363D,2007A&A...468..163D,2005AJ....130.1247L,2008AJ....135.2177L}.}
\tablenotetext{b}{MOTION sample includes $\mu$ $\ge$ 1$\farcs$00 yr$^{-1}$ }
\tablenotetext{c}{SLOWMO sample includes 1$\farcs$00 yr$^{-1}$ $>$ $\mu$ $\ge$ 0$\farcs$50 yr$^{-1}$}
\tablenotetext{d}{MINIMO sample includes 0$\farcs$50 yr$^{-1}$ $>$ $\mu$ $\ge$ 0$\farcs$18 yr$^{-1}$}
\tablenotetext{e}{All SCR searches to date.}
\end{deluxetable}

\clearpage


\begin{deluxetable}{lcccccrrrrrcrl}
\rotate \tabletypesize{\scriptsize} \tablecaption{New SCR Objects With 16.5 $<$ $R_{59F}$ $\le$ 18.0
\label{faintmfive}}
\tablewidth{0pt}

\tablehead{\vspace{-8pt} \\
\colhead{Name}&
\colhead{RA}&
\colhead{DEC}&
\colhead{$\mu$}&
\colhead{$\theta$}&
\colhead{$B_J$}&
\colhead{$R_{59F}$}&
\colhead{$I_{IVN}$}&
\colhead{$J$}&
\colhead{$H$}&
\colhead{$K_s$}&
\colhead{$(R_{59F}-J)$}&
\colhead{Est Dist}&
\colhead{Notes}\\

\colhead{}&
\colhead{(J2000)}&
\colhead{(J2000)}&
\colhead{($\arcsec$ yr$^{-1}$)}&
\colhead{($\degr$)}&
\colhead{}&
\colhead{}&
\colhead{}&
\colhead{}&
\colhead{}&
\colhead{}&
\colhead{}&
\colhead{(pc)}&
\colhead{}}

\startdata
\vspace{0pt} \\

SCR 0001-8110  & 00 01 06.59 & -81 10 38.0 & 0.232 & 103.5   & 18.60  & 16.80  & 14.61  & 13.40  & 12.88  & 12.63  & 3.40 &      118.2 &                  \\
SCR 0001-1037  & 00 01 39.89 & -10 37 51.9 & 0.185 & 171.9   & 19.62  & 17.63  & 15.78  & 14.13  & 13.52  & 13.28  & 3.50 &      136.9 &                  \\
SCR 0001-1744  & 00 01 43.45 & -17 44 49.9 & 0.187 & 093.4   & 19.08  & 17.49  & 15.44  & 13.63  & 13.04  & 12.81  & 3.86 &      104.5 &                  \\
SCR 0001-4236  & 00 01 46.24 & -42 36 23.2 & 0.181 & 108.4   & 20.00  & 17.88  & 17.10  & 16.51  & 15.62  & 16.25  & 1.37 &   [1148.7] & \tablenotemark{a}\\
SCR 0005-0159  & 00 05 25.63 & -01 59 58.7 & 0.182 & 194.0   & 18.94  & 16.89  & 14.77  & 13.44  & 12.88  & 12.60  & 3.45 &      104.6 &                  \\

\enddata

\tablenotetext{a}{Cool subdwarf candidate - unreliable distance.}


\end{deluxetable}

\clearpage


\begin{deluxetable}{lccccccccccccl}
\rotate \tabletypesize{\scriptsize} \tablecaption{New 25 pc Red Dwarf
  Candidates
\label{25}}
\tablewidth{0pt}

\tablehead{\vspace{-8pt} \\
\colhead{Name}&
\colhead{RA}&
\colhead{DEC}&
\colhead{$\mu$}&
\colhead{$\theta$}&
\colhead{$B_J$}&
\colhead{$R_{59F}$}&
\colhead{$I_{IVN}$}&
\colhead{$J$}&
\colhead{$H$}&
\colhead{$K_s$}&
\colhead{$(R_{59F}-J)$}&
\colhead{Est Dist}&
\colhead{Notes}\\

\colhead{}&
\colhead{(J2000)}&
\colhead{(J2000)}&
\colhead{($\arcsec$ yr$^{-1}$)}&
\colhead{($\degr$)}&
\colhead{}&
\colhead{}&
\colhead{}&
\colhead{}&
\colhead{}&
\colhead{}&
\colhead{}&
\colhead{(pc)}&
\colhead{}}

\startdata
\vspace{0pt} \\

SCR 1144-4302  & 11 44 24.88 & -43 02 53.5 & 0.311 & 293.6   & 19.60  & 17.50  & 14.39  & 12.17  & 11.56  & 11.20  & 5.33  &  24.7 &    \\
SCR 1546-5534  & 15 46 41.84 & -55 34 47.0 & 0.433 & 226.8   & 18.99  & 17.22  & 14.03  & 10.21  &  9.55  &  9.11  & 7.01  &   6.7 & \tablenotemark{a}    \\
SCR 1609-3431  & 16 09 46.19 & -34 31 06.9 & 0.249 & 198.5   & 18.48  & 16.61  & 13.99  & 11.58  & 10.99  & 10.67  & 5.03  &  22.5 &    \\

\enddata

\tablenotetext{a}{Distance suspect. See \S~\ref{sec:individual}.}

\end{deluxetable}

\clearpage


\begin{deluxetable}{lcccccccccccl}
\rotate \tabletypesize{\scriptsize} \tablecaption{New SCR White Dwarf Candidates
\label{wd}}
\tablewidth{0pt}

\tablehead{\vspace{-8pt} \\
\colhead{Name}&
\colhead{RA}&
\colhead{DEC}&
\colhead{$\mu$}&
\colhead{$\theta$}&
\colhead{$B_J$}&
\colhead{$R_{59F}$}&
\colhead{$I_{IVN}$}&
\colhead{$J$}&
\colhead{$H$}&
\colhead{$K_s$}&
\colhead{Est. Dist. \tablenotemark{a}}&
\colhead{Notes}\\

\colhead{}&
\colhead{(J2000)}&
\colhead{(J2000)}&
\colhead{($\arcsec$ yr$^{-1}$)}&
\colhead{($\degr$)}&
\colhead{}&
\colhead{}&
\colhead{}&
\colhead{}&
\colhead{}&
\colhead{}&
\colhead{(pc)}&
\colhead{}}

\startdata
\vspace{0pt} \\

\vspace{-11pt} \\
\multicolumn{11}{c}{Candidates Selected Using $(R_{59F}-J)$ RPM Diagram}\\
\tableline\\

SCR 0058-6905  & 00 58 43.53 & -69 05 41.7 & 0.331 & 120.4 & 19.78 & 17.20 & 16.70 & 15.94 & 15.49 & 15.65 & 12.1  &                                \\
SCR 0128-7401  & 01 28 52.26 & -74 01 53.6 & 0.209 & 089.8 & 19.30 & 16.97 & 16.25 & 15.87 & 15.29 & 15.16 & 12.8  &                                \\
SCR 0153-0730B & 01 53 12.29 & -07 30 30.6 & 0.465 & 150.0 & 19.73 & 17.89 & 17.20 & 16.32 & 15.70 & 15.56 & 28.3  & \tablenotemark{b}              \\
SCR 0205-7941  & 02 05 24.81 & -79 41 03.7 & 0.194 & 098.0 & 17.21 & 17.06 & 16.91 & 16.65 & 16.31 & 15.74 & 65.5  &                                \\
SCR 0221-0143  & 02 21 14.88 & -01 43 40.4 & 0.323 & 219.1 & 17.49 & 16.94 & 16.80 & 16.82 & 16.40 & 16.38 & 46.5  &                                \\
SCR 0243-6034  & 02 43 00.34 & -60 34 14.5 & 0.245 & 178.0 & 17.80 & 16.77 & 16.32 & 16.27 & 15.89 & 15.37 & 30.4  &                                \\
SCR 0345-8652  & 03 45 53.24 & -86 52 31.1 & 0.183 & 012.7 & 17.28 & 17.01 & 16.93 & 16.39 & 16.32 & 15.27 & 59.2  &                                \\
SCR 0616-4122  & 06 16 22.05 & -41 22 31.4 & 0.181 & 348.3 & 17.42 & 16.97 & 16.70 & 16.47 & 16.11 & 16.04 & 50.8  &                                \\
SCR 0626-3752  & 06 26 00.05 & -37 52 18.2 & 0.293 & 138.6 & 18.47 & 17.59 & 17.22 & 16.44 & 16.30 & 16.33 & 49.8  &                                \\
SCR 0637-7410  & 06 37 24.25 & -74 10 59.3 & 0.233 & 025.4 & 17.56 & 17.10 & 16.87 & 16.50 & 16.11 & 15.71 & 53.6  &                                \\
SCR 0712-0428  & 07 12 06.17 & -04 28 16.0 & 0.218 & 341.2 & 18.19 & 17.21 & 16.92 & 16.30 & 16.12 & 15.53 & 38.8  &                                \\
SCR 0816-4641  & 08 16 30.13 & -46 41 13.2 & 0.441 & 060.7 & 18.27 & 16.88 & 16.18 & 15.68 & 15.59 & 14.91 & 24.7  &                                \\
SCR 0825-5107  & 08 25 33.14 & -51 07 30.9 & 0.332 & 301.7 & 17.73 & 16.86 & 16.66 & 15.68 & 15.34 & 15.25 & 35.5  &                                \\
SCR 0855-0833  & 08 55 34.71 & -08 33 45.0 & 0.467 & 188.9 & 18.79 & 17.31 & 16.77 & 16.08 & 15.84 & 15.28 & 28.0  &                                \\
SCR 0913-3910B & 09 13 12.40 & -39 10 19.0 & 0.232 & 248.2 & 17.61 & 17.11 & 16.68 & 16.09 & 17.48 & 15.40 & 52.6  & \tablenotemark{b}              \\
SCR 0913-6851  & 09 13 32.27 & -68 51 36.5 & 0.344 & 320.2 & 17.91 & 17.35 & 16.93 & 16.55 & 15.93 & 16.07 & 56.1  &                                \\
SCR 1016-3616  & 10 16 22.15 & -36 16 44.4 & 0.202 & 327.6 & 17.42 & 16.95 & 16.73 & 16.50 & 16.34 & 15.37 & 49.3  &                                \\
SCR 1126-0931  & 11 26 34.26 & -09 31 15.2 & 0.283 & 241.4 & 20.27 & 17.99 & 17.18 & 16.34 & 15.59 & 15.43 & 21.4  &                                \\
SCR 1129-7823  & 11 29 24.40 & -78 23 57.7 & 0.211 & 264.7 & 17.48 & 17.35 & 17.20 & 16.46 & 16.14 & 15.83 & 76.0  &                                \\
SCR 1156-3655  & 11 56 04.30 & -36 55 48.0 & 0.207 & 269.6 & 18.45 & 17.67 & 17.30 & 16.34 & 16.95 & 15.60 & 55.2  &                                \\
SCR 1203-3621  & 12 03 34.26 & -36 21 05.1 & 0.259 & 300.2 & 17.28 & 16.89 & 16.59 & 16.19 & 15.98 & 15.74 & 50.9  &                                \\
SCR 1204-3514  & 12 04 48.89 & -35 14 27.9 & 0.242 & 261.9 & 17.69 & 17.31 & 16.88 & 16.65 & 15.85 & 16.76 & 62.2  &                                \\
SCR 1216-3758  & 12 16 16.94 & -37 58 48.0 & 0.275 & 191.6 & 19.37 & 17.90 & 17.41 & 16.66 & 16.50 & 15.65 & 37.1  &                                \\
SCR 1224-0018  & 12 24 22.91 & -00 18 18.6 & 0.208 & 271.9 & 17.60 & 16.88 & 16.42 & 16.57 & 16.35 & 15.51 & 40.0  &                                \\
SCR 1241-0733  & 12 41 40.12 & -07 33 06.0 & 0.182 & 280.5 & 17.00 & 16.83 & 16.72 & 16.63 & 17.66 & 16.28 & 58.6  &                                \\
SCR 1243-3926  & 12 43 38.77 & -39 26 46.1 & 0.185 & 209.8 & 17.58 & 17.15 & 16.83 & 16.64 & 16.34 & 15.66 & 56.1  &                                \\
SCR 1351-3912  & 13 51 09.73 & -39 12 51.4 & 0.282 & 139.6 & 17.67 & 16.65 & 16.28 & 15.98 & 15.66 & 15.84 & 29.0  &                                \\
SCR 1416-6531  & 14 16 22.46 & -65 31 26.8 & 0.337 & 258.8 & 17.00 & 16.67 & 16.67 & 16.16 & 16.09 & 15.57 & 48.1  &                                \\
SCR 1452-0011  & 14 52 24.95 & -00 11 34.9 & 0.288 & 218.1 & 19.48 & 17.38 & 16.32 & 16.58 & 15.74 & 15.47 & 18.3  &                                \\
SCR 1600-1654  & 16 00 41.16 & -16 54 30.2 & 0.812 & 284.0 & 19.30 & 17.84 & 17.22 & 16.56 & 16.27 & 16.57 & 36.4  &                                \\
SCR 1601-3832  & 16 01 37.02 & -38 32 09.2 & 0.326 & 124.5 & 18.01 & 17.08 & 16.80 & 15.89 & 15.86 & 15.59 & 37.5  &                                \\
SCR 1751-7749B & 17 51 39.55 & -77 49 02.6 & 0.341 & 188.0 & 17.43 & 16.93 & 16.70 & 16.32 & 15.70 & 15.37 & 48.0  & \tablenotemark{b}              \\
SCR 2008-8051  & 20 08 14.23 & -80 51 08.4 & 0.214 & 180.8 & 18.16 & 17.79 & 17.57 & 16.66 & 17.07 & 15.82 & 78.8  &                                \\
SCR 2045-1411  & 20 45 43.03 & -14 11 31.7 & 0.831 & 170.6 & 20.24 & 17.39 & 15.99 & 15.11 & 14.57 & 14.57 & 10.7  &                                \\
SCR 2101-3509  & 21 01 26.68 & -35 09 33.4 & 0.184 & 250.9 & 17.10 & 16.59 & 16.21 & 16.26 & 16.09 & 15.50 & 40.6  &                                \\
SCR 2105-3247  & 21 05 24.18 & -32 47 32.7 & 0.193 & 257.2 & 16.93 & 16.55 & 16.29 & 15.81 & 15.53 & 15.67 & 44.1  &                                \\
SCR 2108-0312  & 21 08 54.88 & -03 12 03.1 & 0.192 & 187.6 & 17.82 & 16.84 & 16.29 & 15.98 & 15.93 & 15.96 & 32.4  &                                \\
SCR 2110-1158  & 21 10 11.80 & -11 58 33.2 & 0.262 & 106.2 & 17.68 & 16.63 & 16.06 & 15.58 & 15.16 & 15.23 & 27.9  &                                \\
SCR 2140-5124  & 21 40 44.77 & -51 24 14.5 & 0.190 & 188.3 & 17.80 & 17.23 & 16.97 & 16.64 & 17.54 & 16.96 & 52.7  &                                \\
SCR 2153-0214  & 21 53 42.11 & -02 14 26.6 & 0.277 & 092.3 & 17.58 & 17.10 & 16.86 & 16.64 & 16.04 & 17.06 & 53.1  &                                \\
SCR 2225-2007  & 22 25 48.57 & -20 07 31.1 & 0.277 & 114.8 & 17.81 & 17.05 & 16.56 & 16.30 & 16.12 & 15.88 & 41.8  &                                \\
SCR 2331-4907  & 23 31 55.99 & -49 07 42.2 & 0.217 & 153.7 & 19.86 & 17.66 & 17.11 & 16.29 & 15.41 & 15.78 & 19.6  &                                \\

\tableline\\
\vspace{-15pt}\\
\multicolumn{11}{c}{Candidates Selected Using $(B-R_{59F})$ RPM Diagram \tablenotemark{c}}\\
\vspace{-8pt}\\
\tableline\\                                                                                                                  

SCR 0019-1113  & 00 19 42.07 & -11 13 58.0 & 0.246 & 112.4  & 18.05 & 17.68 & 17.39 & \nodata& \nodata& \nodata& 74.6   &               \\
SCR 0055-1504  & 00 55 46.25 & -15 04 51.0 & 0.278 & 101.2  & 17.42 & 17.26 & 17.35 & \nodata& \nodata& \nodata& 71.6   &               \\
SCR 0117-1659  & 01 17 52.23 & -16 59 26.0 & 0.198 & 248.2  & 18.67 & 17.75 & 17.33 & \nodata& \nodata& \nodata& 51.8   &               \\
SCR 0120-7851  & 01 20 15.88 & -78 51 43.0 & 0.180 & 166.1  & 16.83 & 16.86 & 16.82 & \nodata& \nodata& \nodata& 68.7   &               \\
SCR 0148-1824  & 01 48 31.64 & -18 24 55.0 & 0.193 & 119.0  & 18.46 & 17.79 & 17.60 & \nodata& \nodata& \nodata& 63.2   &               \\
SCR 0211-1434  & 02 11 51.89 & -14 34 48.0 & 0.197 & 096.4  & 17.47 & 17.32 & 17.37 & \nodata& \nodata& \nodata& 74.2   &               \\
SCR 0220-6936  & 02 20 48.49 & -69 36 32.0 & 0.267 & 079.6  & 18.71 & 17.84 & 17.41 & \nodata& \nodata& \nodata& 56.2   &               \\
SCR 0226-6200  & 02 26 55.97 & -62 00 38.0 & 0.183 & 054.6  & 17.39 & 17.30 & 17.13 & \nodata& \nodata& \nodata& 77.0   &               \\
SCR 0256-0700  & 02 56 41.59 & -07 00 33.0 & 0.423 & 118.1  & 19.06 & 17.79 & 17.39 & \nodata& \nodata& \nodata& 40.6   &               \\
SCR 0321-2930  & 03 21 59.47 & -29 30 33.0 & 0.274 & 082.4  & 18.83 & 17.89 & 17.80 & \nodata& \nodata& \nodata& 54.2   &               \\
SCR 0331-4118  & 03 31 57.25 & -41 18 06.0 & 0.213 & 208.8  & 18.50 & 17.57 & 17.24 & \nodata& \nodata& \nodata& 47.2   &               \\
SCR 0353-6514  & 03 53 33.20 & -65 14 18.0 & 0.187 & 027.3  & 18.08 & 17.65 & 17.67 & \nodata& \nodata& \nodata& 70.9   &               \\
SCR 0401-7643  & 04 01 31.44 & -76 43 52.0 & 0.231 & 025.5  & 18.71 & 17.88 & 17.48 & \nodata& \nodata& \nodata& 58.3   &               \\
SCR 0411-1353  & 04 11 51.39 & -13 53 55.0 & 0.201 & 179.9  & 18.42 & 17.99 & 18.00 & \nodata& \nodata& \nodata& 82.6   &               \\
SCR 0506-5510  & 05 06 40.16 & -55 10 52.0 & 0.182 & 035.0  & 17.42 & 17.16 & 16.95 & \nodata& \nodata& \nodata& 63.7   &               \\
SCR 0632-6503  & 06 32 24.12 & -65 03 59.0 & 0.236 & 014.5  & 18.38 & 17.88 & 17.79 & \nodata& \nodata& \nodata& 74.5   &               \\
SCR 0737-6347  & 07 37 30.65 & -63 47 34.0 & 0.287 & 310.1  & 18.27 & 17.67 & 17.55 & \nodata& \nodata& \nodata& 62.7   &               \\
SCR 0805-1702  & 08 05 03.23 & -17 02 18.0 & 0.420 & 295.2  & 17.59 & 17.48 & 17.67 & \nodata& \nodata& \nodata& 82.0   &               \\
SCR 0809-1234  & 08 09 36.45 & -12 34 03.0 & 0.402 & 150.2  & 19.22 & 17.96 & 17.42 & \nodata& \nodata& \nodata& 44.8   &               \\
SCR 0815-7344  & 08 15 35.63 & -73 44 31.0 & 0.186 & 349.0  & 18.86 & 17.95 & 17.34 & \nodata& \nodata& \nodata& 56.5   &               \\
SCR 0836-3318  & 08 36 38.24 & -33 18 05.0 & 0.272 & 131.2  & 18.21 & 17.65 & 17.50 & \nodata& \nodata& \nodata& 63.9   &               \\
SCR 0853-6733  & 08 53 47.44 & -67 33 32.0 & 0.231 & 336.6  & 18.60 & 17.91 & 17.69 & \nodata& \nodata& \nodata& 66.0   &               \\
SCR 0859-1700  & 08 59 16.81 & -17 00 55.0 & 0.180 & 311.4  & 17.18 & 17.15 & 17.21 & \nodata& \nodata& \nodata& 75.0   &               \\
SCR 0859-6925  & 08 59 24.18 & -69 25 06.0 & 0.295 & 291.2  & 18.52 & 17.83 & 17.40 & \nodata& \nodata& \nodata& 63.1   &               \\
SCR 0914-1534  & 09 14 04.90 & -15 34 06.0 & 0.233 & 166.2  & 18.71 & 17.77 & 17.40 & \nodata& \nodata& \nodata& 51.2   &               \\
SCR 0930-3944  & 09 30 50.87 & -39 44 04.0 & 0.182 & 145.2  & 17.60 & 17.55 & 17.71 & \nodata& \nodata& \nodata& 88.5   &               \\
SCR 1020-0436  & 10 20 55.99 & -04 36 24.0 & 0.202 & 164.9  & 18.42 & 17.97 & 17.68 & \nodata& \nodata& \nodata& 80.8   &               \\
SCR 1037-0818  & 10 37 37.77 & -08 18 22.0 & 0.187 & 250.0  & 18.28 & 17.75 & 17.57 & \nodata& \nodata& \nodata& 68.5   &               \\
SCR 1103-7109  & 11 03 20.12 & -71 09 58.0 & 0.195 & 271.2  & 18.36 & 17.83 & 17.53 & \nodata& \nodata& \nodata& 71.4   &               \\
SCR 1104-5002  & 11 04 42.01 & -50 02 58.0 & 0.277 & 270.8  & 18.94 & 18.00 & 17.60 & \nodata& \nodata& \nodata& 56.8   &               \\
SCR 1128-0537  & 11 28 04.46 & -05 37 17.0 & 0.328 & 142.2  & 18.35 & 17.50 & 17.32 & \nodata& \nodata& \nodata& 48.4   &               \\
SCR 1139-0846  & 11 39 24.46 & -08 46 39.0 & 0.182 & 298.4  & 17.00 & 16.90 & 16.92 & \nodata& \nodata& \nodata& 63.2   &               \\
SCR 1206-0722  & 12 06 05.68 & -07 22 28.0 & 0.240 & 211.1  & 18.86 & 17.77 & 17.27 & \nodata& \nodata& \nodata& 46.4   &               \\
SCR 1227-4558  & 12 27 33.08 & -45 58 00.0 & 0.186 & 290.0  & 18.94 & 17.97 & 17.50 & \nodata& \nodata& \nodata& 54.8   &               \\
SCR 1314-0438  & 13 14 24.47 & -04 38 12.0 & 0.201 & 239.4  & 18.05 & 17.40 & 17.10 & \nodata& \nodata& \nodata& 53.3   &               \\
SCR 1316-3633  & 13 16 05.10 & -36 33 07.0 & 0.207 & 143.8  & 16.92 & 16.95 & 17.10 & \nodata& \nodata& \nodata& 71.3   &               \\
SCR 1331-7743  & 13 31 44.96 & -77 43 02.0 & 0.337 & 247.6  & 18.33 & 17.98 & 17.83 & \nodata& \nodata& \nodata& 86.6   &               \\
SCR 1339-0937  & 13 39 39.29 & -09 37 07.0 & 0.295 & 246.3  & 18.92 & 17.79 & 17.50 & \nodata& \nodata& \nodata& 45.3   &               \\
SCR 1340-1517  & 13 40 01.12 & -15 17 33.0 & 0.450 & 272.4  & 18.69 & 17.70 & 17.32 & \nodata& \nodata& \nodata& 47.7   &               \\
SCR 1347-4848  & 13 47 43.65 & -48 48 47.0 & 0.373 & 218.3  & 18.67 & 17.94 & 17.51 & \nodata& \nodata& \nodata& 64.9   &               \\
SCR 1402-0902  & 14 02 12.89 & -09 02 45.0 & 0.229 & 259.6  & 18.08 & 17.70 & 17.58 & \nodata& \nodata& \nodata& 74.8   &               \\
SCR 1403-1514  & 14 03 42.83 & -15 14 13.0 & 0.392 & 205.8  & 18.81 & 17.86 & 17.45 & \nodata& \nodata& \nodata& 53.1   &               \\
SCR 1418-1157  & 14 18 44.12 & -11 57 14.0 & 0.181 & 263.5  & 18.11 & 17.71 & 17.84 & \nodata& \nodata& \nodata& 74.0   &               \\
SCR 1556-0805  & 15 56 47.31 & -08 05 59.0 & 0.421 & 117.9  & 18.80 & 17.84 & 18.06 & \nodata& \nodata& \nodata& 52.3   &               \\
SCR 1802-4907  & 18 02 58.31 & -49 07 33.0 & 0.284 & 184.0  & 17.05 & 16.72 & 16.64 & \nodata& \nodata& \nodata& 49.3   &               \\
SCR 1848-4619  & 18 48 09.12 & -46 19 56.0 & 0.208 & 196.3  & 17.71 & 17.81 & 17.80 & \nodata& \nodata& \nodata& 112.4  &               \\
SCR 1904-5851  & 19 04 34.30 & -58 51 16.0 & 0.185 & 161.7  & 17.93 & 17.56 & 17.39 & \nodata& \nodata& \nodata& 70.7   &               \\
SCR 1945-3338  & 19 45 21.46 & -33 38 47.0 & 0.213 & 234.2  & 18.64 & 17.99 & 17.88 & \nodata& \nodata& \nodata& 70.3   &               \\
SCR 2049-4237  & 20 49 55.23 & -42 37 35.0 & 0.226 & 126.1  & 18.98 & 17.87 & 17.40 & \nodata& \nodata& \nodata& 47.4   &               \\
SCR 2052-4212  & 20 52 21.29 & -42 12 58.0 & 0.311 & 139.5  & 19.02 & 17.93 & 17.32 & \nodata& \nodata& \nodata& 49.2   &               \\
SCR 2108-1127  & 21 08 37.58 & -11 27 58.0 & 0.228 & 119.3  & 17.08 & 16.99 & 16.99 & \nodata& \nodata& \nodata& 66.9   &               \\
SCR 2115-0741A & 21 15 07.44 & -07 41 33.0 & 0.212 & 187.0  & 17.09 & 17.08 & 16.91 & \nodata& \nodata& \nodata& 73.7   & \tablenotemark{b}\\
SCR 2240-7829  & 22 40 44.71 & -78 29 45.0 & 0.226 & 110.0  & 17.99 & 17.72 & 17.54 & \nodata& \nodata& \nodata& 81.8   &               \\
SCR 2319-0255  & 23 19 35.92 & -02 55 24.0 & 0.216 & 249.2  & 18.37 & 17.55 & 17.33 & \nodata& \nodata& \nodata& 50.8   &               \\
SCR 2334-5111  & 23 34 49.32 & -51 11 51.0 & 0.188 & 095.3  & 17.87 & 17.32 & 17.08 & \nodata& \nodata& \nodata& 55.5   &               \\

\enddata
\tablenotetext{a}{Estimate given using relation of \cite{2001Sci...292..698O}}
\tablenotetext{b}{CPM system - see Table~\ref{cpm}}
\tablenotetext{c}{Candidates lack 2MASS data. Coordinates not J2000.0.}

\end{deluxetable}

\clearpage


\begin{deluxetable}{lcccccccccccll}
\rotate \tabletypesize{\scriptsize} \tablecaption{New SCR Cool
  Subdwarf Candidates
\label{sdfive}}
\tablewidth{0pt}

\tablehead{\vspace{-8pt} \\
\colhead{Name}&
\colhead{RA}&
\colhead{DEC}&
\colhead{$\mu$}&
\colhead{$\theta$}&
\colhead{$B_J$}&
\colhead{$R_{59F}$}&
\colhead{$I_{IVN}$}&
\colhead{$J$}&
\colhead{$H$}&
\colhead{$K_s$}&
\colhead{$(R_{59F}-J)$}&
\colhead{Est. Dist.}&
\colhead{Notes}\\

\colhead{}&
\colhead{(J2000)}&
\colhead{(J2000)}&
\colhead{($\arcsec$ yr$^{-1}$)}&
\colhead{($\degr$)}&
\colhead{}&
\colhead{}&
\colhead{}&
\colhead{}&
\colhead{}&
\colhead{}&
\colhead{}&
\colhead{(pc)}&
\colhead{}}

\startdata
\vspace{0pt} \\

SCR 0001-4236  & 00 01 46.24 & -42 36 23.2 & 0.181 & 108.4 & 19.96  & 17.88  & 17.10  & 16.51  & 15.62  & 16.25  & 1.37    &   [1148.7] &                                               \\
SCR 0008-0028  & 00 08 26.27 & -00 28 50.3 & 0.192 & 177.8 & 18.65  & 16.89  & 16.09  & 15.30  & 14.79  & 14.54  & 1.59    &    [507.9] &                                               \\
SCR 0016-0021  & 00 16 31.20 & -00 21 11.3 & 0.191 & 091.6 & 18.79  & 17.08  & 16.19  & 15.14  & 14.51  & 14.30  & 1.94    &    [432.2] &                                               \\
SCR 0017-5825  & 00 17 45.97 & -58 25 56.0 & 0.195 & 167.3 & 19.89  & 17.79  & 16.65  & 15.43  & 15.11  & 14.88  & 2.36    &    [500.1] &                                               \\
SCR 0017-6821  & 00 17 57.95 & -68 21 53.3 & 0.193 & 114.7 & 19.98  & 17.97  & 16.73  & 15.70  & 15.08  & 14.79  & 2.27    &    [468.1] &                                               \\

\enddata

\end{deluxetable}

\clearpage


\begin{deluxetable}{lrrrlrrrrrl}
\rotate 
\tabletypesize{\scriptsize} 
\tablecaption{Common Proper Motion Systems 
\label{cpm}}
\tablewidth{0pt}
\tablehead{

\colhead{Primary}&
\colhead{$\mu$}&
\colhead{$\theta$}&
\colhead{Distance}&
\colhead{Companion(s)}&
\colhead{$\mu$}&
\colhead{$\theta$}&
\colhead{Distance}&
\colhead{Sep.}&
\colhead{P.A.}&
\colhead{notes}\\

\colhead{}&
\colhead{($\arcsec$ yr$^{-1}$)}&
\colhead{($\degr$)}&
\colhead{(pc)}&
\colhead{} &
\colhead{($\arcsec$ yr$^{-1}$)}&
\colhead{($\degr$)}&
\colhead{(pc)}&
\colhead{($\arcsec$)}&
\colhead{($\degr$)}&
\colhead{}}
\startdata

\vspace{-11pt} \\
\multicolumn{11}{c}{Probable Common Proper Motion Systems}\\
\tableline\\

L 044-086       & 0.318 & 188.7    &    46.7 & SCR 1751-7749B  & 0.341 &  188.0  & [674.2]&  135.3  &     0.4 &   WD candidate at 48.0 pc\tablenotemark{ab}  \\
LEHPM2-2868     & 0.184 & 121.6    &    85.4 & SCR 0253-1738B  & 0.184 &  119.4  &  140.3 &  102.1  &   179.9 &   \tablenotemark{a}                   \\
LEHPM2-5029     & 0.183 & 227.0    &   372.5 & SCR 1312-3103B  & 0.196 &  219.8  &   131.9&  650.7  &   289.4 &                                  \\
SCR 0115-1226A  & 0.192 & 109.4    &   123.9 & SCR 0114-1227B  & 0.196 &  116.2  &   163.9&  464.8  &   186.7 &                                  \\
SCR 0153-0731A  & 0.452 & 140.7    &  [217.8]& SCR 0153-0730B  & 0.465 &  150.0  & [800.1]&   55.5  &   343.2 &   WD candidate at 28.3 pc\tablenotemark{ac}  \\
SCR 0219-7102A  & 0.275 & 041.0    &   137.0 & SCR 0219-7102B  & 0.264 &  043.8  &  205.5 &   34.7  &   275.3 &                            \\
SCR 0307-4654A  & 0.190 & 060.9    &   191.5 & LEHPM1-3098     & 0.194 &  059.1  &   146.9& 1180.3  &    54.7 &                                  \\
SCR 0335-4019A  & 0.188 & 046.7    &    80.6 & SCR 0336-4028B  & 0.184 &  051.7  &   159.9& 1036.6  &   122.0 &                                  \\
SCR 0913-3910A  & 0.233 & 250.0    &    61.4 & SCR 0913-3910B  & 0.232 &  248.2  & [660.5]&   12.8  &   174.2 &   WD candidate at 52.6 pc\tablenotemark{ab} \\
SCR 1018-0130A  & 0.204 & 229.4    & [599.3] & SCR 1017-0143B  & 0.214 &  236.7  & [430.0]& 1039.1  &   232.3 & \tablenotemark{c}                \\
SCR 1352-0205A  & 0.226 & 266.6    &   234.5 & SCR 1352-0157B  & 0.224 &  266.4  &   121.5&  548.9  &    61.2 &                                  \\
SCR 2211-4825A  & 0.183 & 116.0    &   161.3 & SCR 2210-4835B  & 0.181 &  116.9  &   140.0& 1143.0  &   212.5 &                                  \\

\tableline\\
\vspace{-15pt}\\
\multicolumn{11}{c}{Possible Common Proper Motion Systems}\\
\vspace{-8pt}\\
\tableline\\                                                                                                                  

SCR 0203-0727A  & 0.248 & 169.2    &  [677.8]& SCR 0203-0727B  &\nodata& \nodata & \nodata&   15.3  &   339.3 &   \tablenotemark{bc}        \\
SCR 0424-2427A  & 0.181 & 094.5    &   126.2 & LEHPM2-5525     & 0.180 &  128.6  &  112.6 &  183.1  &    29.4 &                            \\
SCR 0443-5732A  &\nodata&\nodata   &  \nodata& SCR 0443-5731B  & 0.259 &  063.8  & [697.9]&   32.6  &    14.7 &   \tablenotemark{c}        \\
SCR 0728-2211A  &\nodata&\nodata   &  \nodata& SCR 0728-2211B  & 0.263 &  181.1  &   99.5 &   20.7  &    13.8 &   \tablenotemark{b}        \\
SCR 0729-8402A  &\nodata&\nodata   &  \nodata& SCR 0729-8402B  & 0.270 &  350.6  & [489.6]&  456.6  &    94.8 &   \tablenotemark{bc}                  \\
SCR 1051-0008A  & 0.269 & 180.0    &    56.5 & SCR 1051-0009B  &\nodata& \nodata & \nodata& \nodata & \nodata &   \tablenotemark{bd} \\
SCR 1351-0115A  & 0.218 & 278.0    & [234.6] & SCR 1351-0118B  & 0.188 &  267.4  &   231.8&  221.0  &   224.2 & \tablenotemark{c}                \\
                &       &          &         & SCR 1352-0106C  & 0.205 &  276.9  &   247.7&  879.6  &    38.2 & \tablenotemark{c}                \\
SCR 1418-4725A  & 0.185 & 191.8    &    69.2 & SCR 1418-4725B  &\nodata& \nodata & \nodata& \nodata & \nodata &   \tablenotemark{bd} \\
SCR 1602-0225A  & 0.189 & 220.7    &   104.0 & SCR 1602-0225B  & 0.184 &  205.1  &   76.4 &   32.9  &   215.4 &                                               \\
SCR 1612-4829A  &\nodata&\nodata   &  \nodata& SCR 1612-4830B  & 0.194 &  234.0  &   71.9 &  161.6  &   212.9 &   \tablenotemark{b}        \\
SCR 2115-0741A  & 0.212 & 187.0    &  \nodata& SCR 2115-0744B  & 0.235 &  160.5  &  140.0 & \nodata & \nodata &   WD candidate at 73.7 pc\tablenotemark{abd} \\
SCR 2236-6030A  &\nodata&\nodata   &  \nodata& SCR 2236-6029B  & 0.260 &  125.4  & [320.4]&   80.3  &    20.9 &   \tablenotemark{bc} \\

\enddata

\tablenotetext{a}{WD candidate with unreliable distance estimate [in brackets]. More accurate estimate in notes if available. See Table~\ref{wd}.}
\tablenotetext{b}{Companion detected by eye during blinking process.}
\tablenotetext{c}{Cool subdwarf candidate with unreliable distance estimate [in brackets]. See Table~\ref{sdfive}.}

\end{deluxetable}

\clearpage


\end{document}